\documentclass[fleqn,12pt,twoside]{article}
\usepackage{espcrc1,amssymb}

\usepackage{graphicx}
\usepackage[figuresright]{rotating}

\newcommand{\AmS}{{\protect\the\textfont2
  A\kern-.1667em\lower.5ex\hbox{M}\kern-.125emS}}

\title{Quark correlations and gluon propagators in elastic vector
meson production\thanks{Work supported by the EC--IHP Network ESOP, 
Contract HPRN-CT-2000-00130.}}

\author{F. Cano\address[CEA]{DAPNIA/SPhN, CEA--Saclay, F91191
Gif-sur-Yvette Cedex, France} and J.-M. Laget\addressmark[CEA]}

\begin{document}

\maketitle

\begin{abstract}
We study the behavior of the differential cross section for vector
meson photoproduction at large momentum transfer in the two--gluon
exchange model. We focus on the treatment of two--quark correlation
function in the proton and on gluon propagators with a dynamically
generated mass. We find that only the large $t$ region is sensitive to
the particular details of these inputs.
\end{abstract}

\section{INTRODUCTION}

	The exchange of two--gluons is the simplest realization of the
	pomeron in terms of the QCD degrees of freedom
	\cite{DONNACHIE89}. It was recently shown that the two--gluon
	exchange model can account for experimental data for the
	differential cross section of $\rho$ and $\phi$
	photoproduction up to a momentum transfer of $-t \lesssim 6$
	GeV$^2$ and $-t \lesssim 3$ GeV$^2$ respectively
	\cite{LAGET00,ANCIANT00}, which is the largest momentum range
	available so far. One of the essential issues in getting such
	a good agreement was the role of quark correlations, i.e.,
	diagrams where each gluon couples to a different quark in the
	proton.
 
	As new data will become available from JLab for a wider range
	of $t$ it is important to study the predictions of the
	two--gluon exchange model in that region. In particular, our
	goal is to test the stability of these predictions against
	changes in the inputs of the model. We will focus on the study
	of the treatment of quark correlations in the proton and on
	the choice of the (non--perturbative) gluon propagator. 

\section{TWO--QUARK CORRELATION FUNCTION}

It was argued that contributions from the diagram of Figure 1b, not
included in the pomeron picture, are negligible as compared to the
ones where both gluons couple to the same quark \cite{LANDSHOFF87},
Figure 1a. While this is true at low $t$, as $t$ increases both
contributions become comparable and eventually the one of Figure 1b
dominates \cite{LAGET00}.

	At high energies the Dirac structure of the two--gluon
	coupling in Figure 1a is simply vector--like. Its contribution
	is proportional to the isoscalar Dirac form factor of the
	nucleon $F_1(-t)$, which is usually taken from fits to the
	experimental data. The diagram of Figure 1b is governed by a
	two--quark correlation function which, in the eikonal
	approximation \cite{GUNION77}, can be written as:
 
\begin{equation}
G_2(\vec{k}_a, \vec{k}_b) = \int \Pi_i d x_i d \vec{r}_i \delta(\sum_i
x_i 
\vec{r}_i) \delta(\sum_i x_i -1) |\Psi(x_i,\vec{r}_i)|^2 
e^{i (\vec{k}_a \cdot \vec{r}_j + \vec{k}_b \cdot \vec{r}_k) } \; ,
\end{equation}

\noindent where $\vec{k}_a$, $\vec{k}_b$ are the (transverse) momenta
flowing through each gluon ($(\vec{k}_a + \vec{k}_b)^2 = -t$) and
$\Psi$ is the nucleon wave function which depend on the longitudinal
momentum $x_i$ and the coordinates $\vec{r}_i$ in the transverse
space. The evaluation of $G_2$ requires the explicit knowledge of the
nucleon wave function. Nonetheless, this requirement is usually
circumvented by making the approximation \cite{CUDELL94}:

\begin{equation}
G_2(\vec{k}_a,\vec{k}_b) = F_1 (\vec{k}_a^2+ \vec{k}_b^2 - \vec{k}_a
\cdot \vec{k}_b ). 
\label{approximation}
\end{equation}
  
However, the expression above is strictly valid only when $x_i=1/3$.
A more accurate approach that accounts for the smearing in the
longitudinal momentum of the quarks obviously involves the explicit
calculation of $G_2$ with a definite choice for the nucleon wave
function. We have taken a simplified version of the wave function
proposed in Ref. \cite{BOLZ96} and extensively used to evaluate (soft)
contributions to a number of observables \cite{DIEHL99}. In momentum
space it can be written as:

\begin{equation}
\Psi(x_i, \vec{k}_i) = N_\Psi \phi_{{\mbox  \small AS}} 
\frac{1}{x_1 x_2 x_3} \exp[- a_N^2 \sum_{i} \frac{\vec{k}_i^2}{x_i}]
\label{wavefunction}
\end{equation}

\noindent where $N_{\Psi}$ is a normalization constant, $\phi_{{\mbox
\small AS}}=120 x_1 x_2 x_3 $ is the asymptotic distribution amplitude
for the nucleon and $\vec{k}_i$ refers to transverse momentum. We have
fitted the free parameter $a_N$ to get a reasonable description of the
isoscalar Dirac form factor of the nucleon. We get $a_N= 1.25$
GeV$^{-1}$, which gives and averaged transverse momentum $\langle
\vec{k}_i^2 \rangle^{1/2} = 0.325 $ GeV.

\begin{figure}
\centerline{\includegraphics[scale=0.6]{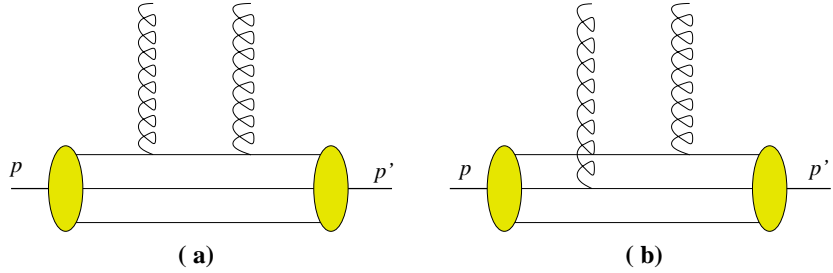}}
\vspace{-1cm}
\caption{Coupling of the gluons to the valence quarks of the proton.}
\end{figure}

\section{GLUON PROPAGATORS}

	Another important component in the two--gluon exchange model
	is the choice of the gluon propagator. In order to get an
	infrared safe behavior one has to deal with dressed
	propagators which are finite at the origin. From the physical
	viewpoint this prevents the gluon from propagating over very
	large distances. In order to match the successful description
	of cross sections provided by the pomeron the two--gluon
	exchange $qq \rightarrow qq$ amplitude is normalized at $t=0$
	to the pomeron exchange amplitude, i.e.

\begin{equation}
\int_0^{\infty} dl^2 [\alpha_n D(l^2)]^2 = \frac{9}{4 \pi} \beta_0^2 \; ,
\label{normalization}
\end{equation}

\noindent where $\beta_0 = 2$ GeV$^{-1}$ is the pomeron--quark
coupling constant and $\alpha_n$ is an effective (frozen) value for
the strong coupling constant which takes into account that we also
deal with the non--perturbative domain.

	It is customary to choose a Gaussian form for the gluon propagator:

\begin{equation}
\alpha_n D(l^2) = \frac{3 \beta_0}{\sqrt{2 \pi} \lambda_0} 
\exp [-l^2/\lambda_0^2]\; ,
\label{gaussian}
\end{equation}
 
\noindent where the parameter $\lambda_0^2=2.7$ GeV$^2$ is fixed to
reproduce the total cross section for $\rho$ electroproduction 
\cite{LAGET95}.  
	
A Gaussian propagator provides a reasonable agreement with a wide
variety of experimental data \cite{LAGET95}. However, it poses a
conceptual problem since it does not have the right asymptotic
behavior. A perturbative tail could be added by hand to
Eq. (\ref{gaussian}), but it can be shown that the (Gaussian)
non--perturbative part dominates the contribution to the differential
cross section even at large $t$. An alternative path is to allow the
gluon to have an effective mass which renders the perturbative
propagator finite at the origin and has the right asymptotic behavior,
provided that this mass vanishes at large momentum. Cornwall
\cite{CORNWALL82} derived an expression for a massive gluon
propagator:

\begin{equation}
\alpha_n D(l^2) = \frac{4 \pi}{9} \frac{1}{[l^2 + m^2(l^2)]
\log{[\frac{l^2 +
4 m^2 (l^2)}{\Lambda^2}]}}
\label{cornwall}
\end{equation}  

\noindent with a dynamically generated mass $m(l^2)$ which has a
logarithmic fall--off with the momentum. In order to set a common
ground to compare with other approaches we have imposed the
normalization condition (\ref{normalization}) on this propagator and
then we find $m(0) = 396$ MeV for $\Lambda= 300$ MeV. This value for
$m(0)$ is within the range of the estimates of Cornwall ($m(0) = 500
\pm 200$ MeV) \cite{CORNWALL82}.


\begin{figure}[t]
\vspace{-1cm}
\begin{center}
\includegraphics[scale=0.35]{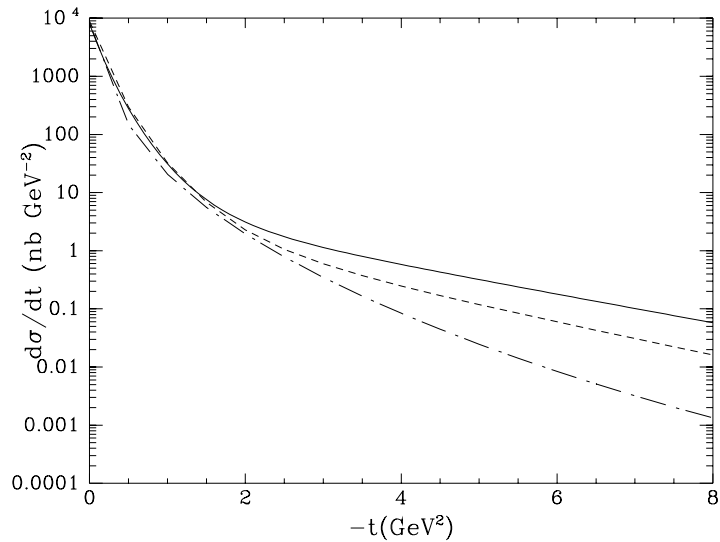} \rule{50pt}{0pt}
\end{center}
\vspace{-1.2cm}
\caption{Differential cross section for $\phi$ photoproduction in the
two--gluon exchange  model (see text for explanation).}
\end{figure}

\section{RESULTS} 
 
In Figure 2 we summarize our results for the differential cross
section for $\phi$ photoproduction. The solid line represents the
calculation with a Gaussian propagator and the assumption
(\ref{approximation}) for the two--quark correlation function. Results
in Ref. \cite{LAGET00} where obtained under these assumptions. If we
explicitly calculate the function $G_2$ with the wave function
(\ref{wavefunction}) then we get the results represented by the dashed
line in Figure 2. By comparing the two curves we can see that the
specific way in which two--quark correlations are calculated has some
effects only at large momentum transfer: at low and moderate $t$ it is
the diagram of figure 1a, i.e. the isoscalar Dirac form factor, that
dominates. Dashed--dotted line is obtained with $G_2$ evaluated with
the nucleon wave function and the massive propagator for the gluon,
Eq. (\ref{cornwall}). By comparing with the dashed line it is clear
that the use of a massive propagator instead of the Gaussian one
produces a further depletion in the differential cross section at
large $t$. It also decreases slightly the slope at small $t$.

The same pattern of differences shown in Figure 2 is obtained for
$\rho$ production. However, in that case, those differences are
softened by quark--exchange contributions, which have to be
incorporated before comparing with data \cite{CANO01}.
  
It is worthwhile emphasizing that preliminary data from the CLAS
Collaboration at JLab for $\phi$ photoproduction at $E_\gamma=4.5$ GeV
supports the depletion obtained with the massive gluon propagator and
the explicit calculation of $G_2$ (dashed--dotted line in Figure
2). Moreover, the use of a massive propagator is also essential in
getting a definite asymptotic behavior for $d\sigma/dt$ at attainable
values of $t$. We will address these issues in more detail in
\cite{CANO01}.


\begin{thebibliography}{9}
\bibitem{DONNACHIE89} A. Donnachie and P.V. Landshoff, 
Nucl. Phys. B 311 (1989) 509.
\bibitem{LAGET00} J.-M. Laget, Phys. Lett. B 489 (2000) 313.
\bibitem{ANCIANT00} E. Anciant et al., Phys. Rev. Lett. 85 (2000)
4682.
\bibitem{LANDSHOFF87} P.V. Landshoff and O. Nachtmann, 
Z. Phys. C 35 (1987) 405.
\bibitem{GUNION77} J.F. Gunion and D.E. Soper, Phys. Rev. D 15 (1977) 2617.
\bibitem{CUDELL94} J.R. Cudell and B.U. Nguyen, Nucl. Phys. B 420 (1994) 669.
\bibitem{BOLZ96} J. Bolz and P. Kroll, Z. Phys. A 356 (1996) 327.
\bibitem{DIEHL99} M. Diehl, Th. Feldmann, R. Jakob and P. Kroll, 
Eur. Phys. J. C 8 (1999) 409.
\bibitem{LAGET95} J.-M. Laget and R. Mendez-Galain, 
Nucl. Phys. A 581 (1995) 397.
\bibitem{CORNWALL82} J.M. Cornwall, Phys. Rev. D 26 (1982) 1453.
\bibitem{CANO01} F. Cano and J.-M. Laget, in preparation.
\end{thebibliography}
\end{document}